\PassOptionsToPackage{usenames,dvipsnames,svgnames}{xcolor}
\documentclass[preprint,journal]{vgtc}            

\onlineid{0}

\vgtccategory{Research}

\vgtcpapertype{please specify}

\title{Challenges and Opportunities in Data Visualization Education:\\ A Call to Action}

\author{%
\authororcid{Benjamin Bach}{0000-0002-9201-7744},
\authororcid{Mandy Keck}{0000-0002-5821-8016},
\authororcid{Fateme Rajabiyazdi}{0000-0002-8710-865X},
\authororcid{Tatiana Losev}{0000-0002-0363-8072},
\authororcid{Isabel Meirelles}{0000-0001-8111-6002},
\authororcid{Jason Dykes}{0000-0002-8096-5763},\\
\authororcid{Robert S. Laramee}{0000-0002-3874-6145},
Mashael AlKadi,
\authororcid{Christina Stoiber}{0000-0002-1764-1467},
\authororcid{Samuel Huron}{0000-0002-9319-8559},
\authororcid{Charles Perin}{0000-0002-7324-9363},\\
Luiz Morais,
\authororcid{Wolfgang Aigner}{0000-0001-5762-1869},
\authororcid{Doris Kosminsky}{0000-0003-2087-3928},
\authororcid{Magdalena Boucher}{0000-0003-3406-8743},
\authororcid{Søren Knudsen}{0000-0002-8306-1102},\\
Areti Manataki,
\authororcid{Jan Aerts}{0000-0002-6416-2717},
\authororcid{Uta Hinrichs}{0000-0001-7494-0941},
  \authororcid{Jonathan~C. Roberts}{0000-0001-7718-3181} and
\authororcid{Sheelagh Carpendale}{0000-0002-5127-9780}
}

\newcommand{\numchallenges}{19}

\newcommand{\numgrandchallenges}{5}

\newcommand{\numquestions}{43}

\newcommand{\new}[2]{#1}

\newcommand{\rqPPLhow}{How do we establish who teaches data visualization, understand how they do it and how they succeed?} 

\newcommand{\rqPPLbck}{How can we understand the effects of an educator's background on teaching philosophy, goals and methods?}

\newcommand{\rqPPLrng}{How can we develop educational offerings that support a wide range of backgrounds, methods, and goals?}

\newcommand{\rqPPLlrn}{How do we describe learners in visualization, their goals, motivations and individual needs?}

\newcommand{\rqPPLdiv}{How can we understand the diversity in learners and what this brings to their learning and experience?}

\newcommand{\rqPPLent}{How do we develop entry points to visualization education that embrace learners' diversity?}

\newcommand{\rqPPLflx}{How do we work with learners to co-design goals, methods, and learning pathways?}

\newcommand{\rqPPLcol}{How do we develop ways in which educators and learners learn from each other?}

\newcommand{\rqPPLper}{How can we develop ways in which learners take an active part in the education of their peers?}

\newcommand{\rqGAgen}{How can we develop learning goals specific to data visualization, including for diverse groups of learners?}

\newcommand{\rqGAgrp}{How to evaluate assessment methods for these learning goals?}

\newcommand{\rqGAskl}{How do we assess skills related to creativity, problem-solving, collaboration, and group work in data visualization?}

\newcommand{\rqGAdiv}{How do we assess learning in diverse, varied, informal, and work-based education scenarios?}

\newcommand{\rqGAdif}{How do we assess learners working with different (self-selected, personal) datasets, technologies, collaborators or organizations?}

\newcommand{\rqGAfbk}{How do we provide feedback that is transparent, fair, high-quality, and timely when teaching at scale or distance?}

\newcommand{\rqMTVdiv}{How to better communicate goals, levels, expectations, and prerequisites to diverse groups of learners?}

\newcommand{\rqMTVfac}{How do we understand and respond to the factors that motivate people to engage with visualization learning?}

\newcommand{\rqMTVsuc}{How do we capture and share visualization success stories and understand what made them a success?}
\newcommand{\rqMTVbck}{How do we embrace learners' disciplinary and background knowledge in ways that motivate?}
\newcommand{\rqMTVflx}{How can we balance flexibility in learning, through free-form, learner-centered and individualized activity, with educator workload?}

\newcommand{\rqMTHskl}{How do we identify, prioritize, and develop core skills in visualization?}

\newcommand{\rqMTHply}{How do we leverage and understand play for data visualization education?}

\newcommand{\rqMTHint}{How do we support the development of skills and knowledge in interactive exploration and the use of sophisticated visual analytics tools and techniques?}

\newcommand{\rqMTHcmb}{How do we combine core skills in visualization with broader competencies?}

\newcommand{\rqMTHfnd}{How can we describe, collect, and share the range of activities and methods available for data visualization education?}
\newcommand{\rqMTHwrk}{How do we substantiate the activities and methods that work well for individual learners, groups of learners, and diverse contexts?}

\newcommand{\rqENVlen}{How do specific learning environments affect data visualization education?}

\newcommand{\rqENVaff}{How do we create approaches that leverage the affordances of specific places, platforms, and other contexts?}

\newcommand{\rqENVtch}{How do we use technology and distributed settings to deliver educational experiences?}

\newcommand{\rqENVdst}{How can we mitigate the limitations of distant learning for social interaction, hands-on activities, and feedback?}

\newcommand{\rqMATfnd}{How can we collect, share, and expand the range of materials and methods for data visualization education?}

\newcommand{\rqMATwrk}{How can we substantiate the materials that work well for individual learners, groups of learners, and diverse contexts?}

\newcommand{\rqMATshr}{How can we share materials for data visualization education in ways that are as accessible and inclusive as possible?}

\newcommand{\rqMATsus}{How can we support, promote and encourage the re-use of materials and move towards environmental sustainability?}

\newcommand{\rqMATfrm}{How can we develop frameworks for sharing and re-using materials for data visualization education?}

\newcommand{\rqMATown}{How can we create, modify and maintain bespoke materials?}

\newcommand{\rqMATgap}{How can we identify and address gaps in data visualization education provision and develop materials that support this activity?}

\newcommand{\rqMATint}{How do we develop intelligent tutoring systems for visualization that provide bespoke feedback and guide learners in their education?}

\newcommand{\rqCHGass}{How do we conduct reliable assessment of learners' competencies when AI can outperform students in the kinds of tasks traditionally valued in data visualization education?}

\newcommand{\rqCHGuse}{How do we determine appropriate use of and needs for data visualization as technology changes the discipline and its practice?}

\newcommand{\rqCHGcmp}{How do we revise the competencies associated with data visualization education as technology changes the discipline and its practice?}  

\newcommand{\rqCHGfut}{How can we develop and engage in futuring activities for data visualization education to identify and react to change?}

\newcommand{\rqCHGrel}{How do we develop and support approaches to education and educational systems that are relevant and robust in light of change?}

\abstract{This paper is a call to action for research and discussion on data visualization education. 
As visualization evolves and spreads through our professional and personal lives, we need to understand how to support and empower a broad and diverse community of learners in visualization. Data Visualization is a diverse and dynamic discipline that combines knowledge from different fields, is tailored to suit diverse audiences and contexts, and frequently incorporates tacit knowledge. This complex nature leads to a series of interrelated challenges for data visualization education.
Driven by a lack of consolidated knowledge, overview, and orientation for visualization education, the 21 authors of this paper---educators and researchers in data visualization---identify and describe \numchallenges{} challenges informed by our collective practical experience.
We organize these challenges around seven themes \textit{People}, \textit{Goals \& Assessment}, \textit{Environment}, \textit{Motivation}, \textit{Methods}, \textit{Materials}, and \textit{Change}. 
Across these themes, we formulate \numquestions{} research questions to address these challenges. As part of our call to action, we then conclude with \numgrandchallenges{} cross-cutting \textit{opportunities} and respective action items: embrace \odiversity, build \ocommunity, conduct \oresearch, act \texttt{AGILE}, and relish \oresponsibility.
We aim to inspire researchers, educators and learners to drive visualization education forward and discuss why, how, who and where we educate, as we learn to 
use visualization to address challenges across many scales and many domains in a rapidly changing world: \url{viseducationchallenges.github.io}.}

\keywords{Data Visualization, Education, Challenges\vspace{-1.5em}
}

\teaser{
  \centering
    \includegraphics[width=1\textwidth]{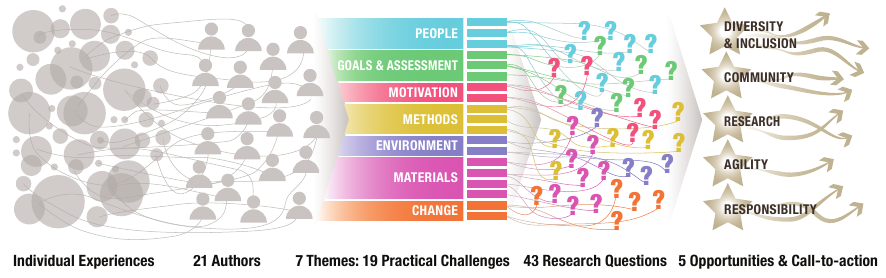}
    \vspace{-2.5em}
    \caption{Overview of contributions, contributors and methods.}
    \label{fig:schema}
    }

\graphicspath{{figs/}{figures/}{pictures/}{images/}{./}} 

\usepackage{tabu}                      
\usepackage{booktabs}                  
\usepackage{lipsum}                    
\usepackage{mwe}                       
\usepackage{color,soul,transparent}
\usepackage{enumitem}

\usepackage{mathptmx}                  

\usepackage{wrapfig}

\newcommand{\minifig}[1]
{
    \begin{wrapfigure}{l}{.6cm}
    \vspace{-0.4cm}
    \includegraphics[width=.8cm]{icons/#1.png}
    \vspace{-0.85cm}
    \end{wrapfigure}
}

\begin{document}


\definecolor{challenge}{HTML}{ffeeaa}
\definecolor{cta}{HTML}{ffebe6}
\definecolor{gc}{HTML}{ffc2b3}
\definecolor{question}{HTML}{e0f5ff}
\definecolor{call}{HTML}{ff00bf}

\newcommand{\chtag}[2]{\textsc{#1#2}}

\newcommand{\challenge}[2]{\noindent\textbf{\underline{#1:} #2}---}

\newcommand{\cta}[1]{\sethlcolor{cta}\hl{#1}}

\newcommand{\gc}[1]{\textbf{#1}}

\newcommand{\secpeople}{\textsc{People}\xspace}
\newcommand{\chpeople}{\textsc{Ppl}}
\newcommand{\chpeopleone}{\chtag{\chpeople}{1}\xspace}
\newcommand{\chpeopletwo}{\chtag{\chpeople}{2}\xspace}
\newcommand{\chpeoplethree}{\chtag{\chpeople}{3}\xspace}
\newcommand{\chpeoplefour}{\chtag{\chpeople}{4}\xspace}
\newcommand{\chgoals}{\textsc{GA}}
\newcommand{\secgoals}{\textsc{Goals\&Assessment}\xspace}
\newcommand{\chgoalsone}{\chtag{\chgoals}{1}\xspace}
\newcommand{\chgoalstwo}{\chtag{\chgoals}{2}\xspace}
\newcommand{\chgoalsthree}{\chtag{\chgoals}{3}\xspace}
\newcommand{\chgoalsfour}{\chtag{\chgoals}{4}\xspace}
\newcommand{\chgoalsfive}{\chtag{\chgoals}{5}\xspace}
\newcommand{\chgoalssix}{\chtag{\chgoals}{6}\xspace}

\newcommand{\chmotivation}{\textsc{Mtv}}
\newcommand{\secmotivation}{\textsc{Motivation}\xspace}
\newcommand{\chmotivationone}{\chtag{\chmotivation}{1}\xspace}
\newcommand{\chmotivationtwo}{\chtag{\chmotivation}{2}\xspace}
\newcommand{\chmotivationthree}{\chtag{\chmotivation}{3}\xspace}
\newcommand{\chmotivationfour}{\chtag{\chmotivation}{4}\xspace}

\newcommand{\chmethods}{\textsc{Mth}}
\newcommand{\secmethods}{\textsc{Methods}\xspace}
\newcommand{\chmethodsone}{\chtag{\chmethods}{1}}
\newcommand{\chmethodstwo}{\chtag{\chmethods}{2}}
\newcommand{\chmethodsthree}{\chtag{\chmethods}{3}}
\newcommand{\chmethodsfour}{\chtag{\chmethods}{4}}

\newcommand{\chresources}{\textsc{Mat}}
\newcommand{\secresources}{\textsc{Materials}\xspace}
\newcommand{\chresourcesone}{\chtag{\chresources}{1}\xspace}
\newcommand{\chresourcestwo}{\chtag{\chresources}{2}\xspace}
\newcommand{\chresourcesthree}{\chtag{\chresources}{3}\xspace}
\newcommand{\chresourcesfour}{\chtag{\chresources}{4}\xspace}
\newcommand{\chresourcesfive}{\chtag{\chresources}{5}\xspace}
\newcommand{\chresourcessix}{\chtag{\chresources}{6}\xspace}
\newcommand{\chresourcesseven}{\chtag{\chresources}{7}\xspace}
\newcommand{\chresourceseight}{\chtag{\chresources}{8}\xspace}

\newcommand{\chenvironment}{\textsc{Env}}
\newcommand{\secenvironment}{\textsc{Environment}\xspace}
\newcommand{\chenvironmentone}{\chtag{\chenvironment}{1}\xspace}
\newcommand{\chenvironmenttwo}{\chtag{\chenvironment}{2}\xspace}

\newcommand{\chchange}{\textsc{Chg}}
\newcommand{\secchange}{\textsc{Change}\xspace}
\newcommand{\chchangeone}{\chtag{\chchange}{1}\xspace}
\newcommand{\chchangetwo}{\chtag{\chchange}{2}\xspace}
\newcommand{\chchangethree}{\chtag{\chchange}{3}\xspace}

\newcommand{\odiversity}{\texttt{DIVERSITY+INCLUSION}}
\newcommand{\oresearch}{\texttt{RESEARCH}}
\newcommand{\ocommunity}{\texttt{COMMUNITIES}}
\newcommand{\oagility}{\texttt{AGILITY}}
\newcommand{\oresponsibility}{\texttt{RESPONSIBILITY}}

\newcounter{questioncounter}
\setcounter{questioncounter}{1}

\newcommand{\question}[1]{
\textcolor{black}{\footnotesize
\textbf{\\Q\arabic{questioncounter}:}}
{\textit{\textcolor{gray}{\scriptsize\textsf{#1}}}}
\stepcounter{questioncounter}
\vspace{0.25mm}
}

\newcommand{\questionformat}[2]{
\textcolor{orange}{\footnotesize#1}
{\textit{\textcolor{gray}{\scriptsize\textsf{#2}}}}
}

\newcommand{\info}[1]{\textcolor{blue}{INFO: \textit{#1}}}

\newcommand{\call}[1]{\textcolor{white}{\sethlcolor{call}\hl{@#1}}}

\newcommand{\ben}[1]{\textit{\footnotesize\textcolor{red}{ben: #1}}}
\newcommand{\uta}[1]{\textit{\footnotesize\textcolor{blue}{uta: #1}}}

\newcommand{\jd}[1]{{\textit{\textcolor{YellowGreen}{jsn: #1}}}} 
\newcommand{\jsn}[1]{{\textit{\textcolor{YellowGreen}{jsn: #1}}}} 
\newcommand{\mandy}[1]{{\textit{\textcolor{YellowGreen}{mandy: #1}}}} 
\newcommand{\tatiana}[1]{\textit{\textcolor{olive}{tatiana: #1}}}
\newcommand{\isabel}[1]{\textit{\textcolor{cyan}{isa: #1}}}
\newcommand{\momo}[1]{\textit{\textcolor{teal}{momo: #1}}}
\newcommand{\wolfgang}[1]{\textit{\textcolor{brown}{wa: #1}}}
\newcommand{\jcr}[1]{{\textit{\textcolor{blue}{jcr: #1}}}} 
\newcommand{\soren}[1]{{\textit{\textcolor{orange}{soren: #1}}}} 
\newcommand{\fateme}[1]{{\textit{\textcolor{gray}{fr: #1}}}}
\newcommand{\charles}[1]{{\textit{\textcolor{blue}{cp: #1}}}}
\newcommand{\review}[1]{{\textit{\textcolor{red}{#1}}}} 

\newcommand{\tag}[1]{{\footnotesize\textsc{[#1]}}} 
\newcommand{\say}[1]{\textit{``#1''}} 

\newcommand{\fix}[1]{\textbf{\textcolor{red}{#1}}}

\firstsection{Introduction}
\maketitle

Education in data visualization is crucial for a data-literate, informed, and critical society that is ever more reliant on data in all spheres of life. As evidence of that, we are seeing increasing numbers of visualizations in news, as part of public dashboards, and in scientific reports~\cite{ipcc2022}. Likewise, we see an increase in the number of textbooks, software tools, courses, workshops, and other resources~\cite{liu:visualization} to aid with the reading, design, and creation of visualizations.
Given these developments and the indisputable power of visualization to analyze, explore, communicate---and equally manipulate and deceive~\cite{pandey2015deceptive}---visualization literacy skills~\cite{boy_principled_2014,borner_data_2019} are paramount.
Consequently, education in visualization becomes a crucial frontier within a discipline that, despite origins millennia back, is in many ways nascent.
As educators and members of a global (scientific) community at the forefront of knowledge generation, it is our responsibility to engage in knowledge transfer, develop creative and practical approaches to education, and lead the research necessary to 
understand and improve education.

\new{While a lot can be learned about education from other disciplines and general educational theories, education for visualization requires individual consideration for a variety of reasons.}{ben} Visualization as a field integrates knowledge and methods from a wide range of disciplines such as statistics, design, art, geography, biology, psychology, cognitive science, computer science, and data science, to name just a few. It is a discipline that combines complex theory and applied craft, diverse domain knowledge and applied problem-solving skills, creativity, and criticality, design expertise and design thinking---all alongside a profound familiarity with data, technology, and human beings. \new{Visualization also creates a growing set of very specific knowledge and methods for, e.g., visualization evaluation~\cite{lam2011empirical}, visualization techniques and algorithms, visual variables and mappings~\cite{carpendale2003considering}, visualization design~\cite{pandey2015deceptive} and design patterns~\cite{bach2022dashboard}, interaction~\cite{yi2007toward}, case studies~\cite{sedlmair2012design}, technology~\cite{ens2021grand}, 
data-driven storytelling~\cite{riche2018data},
task abstraction and theoretic models~\cite{munzner2009nested}, or programming libraries and tools for visualizations~\cite{bostock2011d3}.
Eventually, because of its applied nature and consequent demand across disciplines, visualization education needs to go beyond traditional classroom and student-oriented education and start cater to audiences as diverse as PhD students, children, researchers, working professionals, educators, decision-makers, and domain collaborators, taking into account their respective agendas, goals, and contexts.}{ben}
%
However, despite many dynamic scientific initiatives at  workshops~\cite{WorkshopPedagogy2016,WorkshopPedagogy2017,VisGuides2022,huron2021ieee,huron2020ieee,workshop2006}, panels~\cite{Panel2015,panel2010}, 
and a special issue on visualization literacy~\cite{SpecialIssueVisLiteracy}, visualization is still lacking a
theory \new{of education}{jason} \new{and an established body of knowledge}{ben} of its own. Hence, it is timely to ask questions such as: 
\textit{What are the challenges in visualization education?
What can we do as researchers to address these?
What opportunities arise from this for the visualization community?}

\new{In this paper, we issue a call-to-action for answering these questions and to motivate research in visualization education. Our work initiated from a 1-week Dagstuhl research seminar on \textit{Visualization Empowerment: How to Teach and Learn Data Visualization} in the summer of  2022~\cite{bach2023visualization,bach_et_al:DagstuhlRep2022}. It then evolved over the following year through regular reading, reflection, discussion and knowledge exchange to share, situate and structure our \textbf{personal experiences with visualization education} (\cref{sec:methodology}). This process led us to describe \numchallenges{} challenges grouped into seven themes: \secpeople, \secgoals, \secmotivation, \secmethods, \secenvironment, \secresources, and \secchange (see \cref{sec:challenges}). Our goal is to discuss these challenges in the context of visualization and capture a rich picture of the current state of visualization education as experienced by the 21 authors as educators and academics in data visualization. This bottom-up process aims to provide an empirical reference point to inform discussion, research, and practical advice for visualization educators as well as inspire the community to work towards broader and better data visualization education.}{ben}

To that end, each of our challenges led us to formulate one or more \textit{research questions} (Q1-Q\numquestions) that highlight the need for concerted activity: issues we need to better understand, solutions we need to invent, approaches we need to evaluate, theories we need to create, etc. 
Eventually, our call-to-action details five \textit{opportunities}---%
\odiversity, 
\ocommunity, 
\oresearch, 
\oagility, and 
\oresponsibility{} More info and supplementary material on \url{viseducationchallenges.github.io}.

\section{Background and Related Work}
\label{sec:literature}

Formulating challenges and research agendas is a useful form to reflect and generate new perspectives to facilitate change in the community. 
For example, writings in visualization \textit{research} discuss unsolved challenges in scientific visualization~\cite{johnson2005nihnsf,johnson:top}, immersive analytics~\cite{ens2021grand},
visual analytics~\cite{thomas:illuminating},
information visualization~\cite{chen:top,johnson:future},
human-centered and multifield visualization~\cite{laramee:futureMultiField}, or top challenges for visualization and machine learning~\cite{cagatay:on}. Agendas about \textit{education} elucidate general computing education challenges (e.g., establishing e-learning as a credible alternative to face-to-face) and list specific agenda items~\cite{McGettrick2005}. 
\new{Papers calling for \textit{education in visualization} have a long tradition, too, perhaps starting with Domik asking \textit{``Do We Need Formal Education in Visualization?''} in 2000~\cite{domik2000we}. That article summarizes discussions from the ACM SigGraph subcommittee of Education for Visualization during the 1990s and lists interrelated reasons for formal education, including:
\textit{erroneous interpretations of visualizations are created carelessly};
\textit{substantial amounts of complex knowledge being necessary to prevent this};
and, 
\textit{decision making being increasingly based on visual representations}.
These reasons remain, but the skill sets, competencies and contexts have developed significantly, in particular in the area of information visualization, concerned with more abstract visual representations. Later papers expanded and reinforced Domik's call in 2007~\cite{rushmeier_revisiting_2007}, 2008~\cite{kerren2008teaching}, and 2019~\cite{ryan2019teaching}. However, no one has attempted a broad, experience-based, and structured approach to describing challenges and research questions in visualization education.}{ben}

\new{We are aware that many challenges faced by visualization educators and learners are common to education in general. Instructors need to design teaching situations~\cite{goodyear2015teaching}, 
motivate students to learn~\cite{dewey1938experience}, actively promote the topic and its purpose~\cite{dewey1938experience}, 
deliver accurate content, develop individuals' skills, create an effective and inclusive learning environment~\cite{leifler_teachers_2020}, design and appropriate learning materials~\cite{DIganzio2018}, and account for cultural, cognitive, and disciplinary diversity of learners~\cite{firat:inclusivity,romanelli_learning_2009,tsai_empowering_2020}.}{ben, soren}
\new{We discuss how these challenges unfold in visualization and what that implies for research on 
visualization literacy, design and critical thinking, creativity, ethical grounding, and computational skills.}{jason}

Visual literacy is 
considered
\textit{``the ability to `read', interpret, and understand the information presented in pictorial or graphic images''}~\cite{wileman1993visual} (p. 114). More specifically, visualization literacy can be described as \textit{``the ability to confidently use a given data visualization to translate questions specified in the data domain into visual queries in the visual domain, as well as interpreting visual patterns in the visual domain as properties in the data domain.''} \cite{boy_principled_2014}.
These capabilities relate closely to \emph{graphicacy}
a key competency with persuasive advocates ~\cite{Balchin1976} that sits alongside literacy, numeracy and articulacy as an educational foundation.
Visualization educators should aim to develop these competencies into what Kerzner et al.\ \cite{kerzner2018framework} describe as a ``visualization mindset'', or \textit{``the set of beliefs and attitudes held [...], including an evolving understanding about domain challenges and visualization''}. 

Consequently, many researchers have developed tests for visualization literacy~\cite{lee2016vlat,ge2023calvi,archambault2023mini} and  approaches to help develop skills: from hands-on activities~\cite{huron2021ieee}, learning-by-playing \cite{Amabili2021}, peer reviewing \cite{Friedman2021}, data physicalizations~\cite{wun2016comparing, huron2014constructive,perin2021students}
to broader methods such as design sketching~\cite{roberts2015sketching},  project-based learning \cite{kammer_experience_2021}, and learning through analogy~\cite{ruchikachorn2015learning}. 
\new{Most recently, Firat et al.~\cite{firat:interactive} review over 70 papers that examine, test and study visualization literacy skills. They 
\new{provide inspiration}{soren}
for our work but do not address wider issues associated 
\new{with educational }{soren}
methods, contexts and research that develop literacy.}{Bob, jason}

\section{Methodology}
\label{sec:methodology}

\subsection{Authors}
\label{sec:survey}

We are 21 educators and researchers in academia (10 F, 11 M), teaching in the UK (7), Canada (5), Austria (4), Brazil (2), France (1), Belgium (1), and Denmark (1)---who all met at the Dagstuhl seminar mentioned above. 
\new{Through a structured survey, we collected information about authors’ visualization courses (i.e., topics, methods, tools, and materials) and target audiences. This section summarizes the results of this survey, while more information can be found in the supplementary materials.}{mandy}

\new{Our self-described backgrounds include \textit{Information Visualization, Scientific Visualization, Visual Analytics, Human-Computer Interaction, Epistemology, Design}, and \textit{Cartography}. We work at schools and departments in computer science and engineering (9), art and design (4), and institutes that embrace a multidisciplinary approach (8), including bioinformatics, interactive arts and technology, and communication science. We are early career educators and established faculty with teaching experience from 1-33 years (average: 8 years).}{mandy+fatema}
We teach from university undergraduate levels upwards, including master, PhD, and working professionals.
In addition to 1:1 supervision, we teach classes ranging in size from \new{fewer than 10 students to more than 200.}{Jason}

The topics we teach include the history of data visualization, trends of change, critical and ethical perspectives on data visualization, data and task abstraction, visualization techniques, visual encoding and design, cognition and perception, design processes and guidelines, visualization development, evaluation, and communication with data visualization.
We employ various activities in class, including 
data generation and preparation, sketching, constructing/deconstructing visualizations, card-based games, critiquing visualizations, and
technical learning.
Most of us use a project-based approach and encourage students to bring their own datasets for their projects. Some of us engage students with real-world datasets from open repositories, APIs or external collaborators. 
\new{We use and teach various tools, such as design tools (e.g., Adobe Illustrator), visualization frameworks (e.g., D3.js) or visualization software (e.g., Tableau).}{Mandy}

So, while we represent highly western-centric and, hence, evidently partial, perspectives on visualization education, our experience of visualization education covers a broad range of contexts, approaches, topics, methods, and tools.
This further emphasizes the complexity of this topic, challenges, and the questions that follow.

\subsection{Identifying Challenges, Questions, and Opportunities}

We identified several phases in our process of
characterizing and clarifying
the challenges: preparing, sharing, defining, developing, refining, calling-to-action.
In \textbf{preparing} for 
the Dagstuhl seminar~\cite{bach_et_al:DagstuhlRep2022}, the four seminar organizers invited 66~researchers, practitioners and educators active in the areas of visualization, pedagogy and education, art \& design, and cognitive psychology. The organizers strove for participant diversity in terms of disciplinary background, gender, geographic location, and level of experience. 41 participants accepted the invitation.

\textbf{(1) Sharing experiences} at Dagstuhl helped participants understand the diversity of the attendees, which was achieved through talks, hands-on and brainstorming activities, as well as group discussions. Residential workshops, such as Dagstuhl, enable people to socialize, in formal and informal ways, providing a rich environment for \textbf{open academic discussions and reflection on visualization teaching practices}. For instance, the topic brainstorming session enabled themes, challenges and topics to be confirmed, resulting in the formation of working groups on \textit{Creativity}, \textit{Diversity}, and \textit{Teaching Methods} among others. In each group, practices and experiences were shared and discussed freely and without specific external agenda. While the groups worked independently on their respective questions and outputs, links were made between groups and discussion topics through shared files, backchannel instant messaging, and daily progress presentations.
Discussions within these working groups had a significant impact on the themes, challenges and questions we describe in this paper. To \textbf{(2) collect the challenges}, 15~participants
gathered on the last day of the seminar for a 1h  brainstorm to write an \textbf{initial set of challenges} in education, prompted through their experiences and discussions and reflections at the seminar. A shared document was jointly edited, producing an initial list of approximately  60~individual challenges (which we referenced as LIST-1). Each challenge had a title and a few accompanying sentences.

The next phase was to \textbf{(3) develop, refine, and group} the challenges in LIST-1, beyond the seminar. Through several all-invited teleconference meetings, we started discussions to group challenges into higher-level concepts, which we called \textit{themes}: \textit{Audiences, Content, Methods, Evaluation, Motivation}, and \textit{Resources}. Our idea was to categorize the challenges, to better grasp the range of themes, but also to practically divide into sub-groups and delve into details within each theme.
\new{We found that these themes corresponded to general themes in education (e.g.,~\cite{dewey1938experience,goodyear2015teaching}) giving us confidence that we covered a wide range of topics.}{ben, soren, jason SUMMARY1}
We discussed alternative groupings that, e.g., differentiate between challenges for educators and challenges for learners, research-related vs. practical challenges, and short- and long-term challenges.  These groupings did not materialize further as we found them to be either too high-level and unspecific or too strongly linked and narrow. 
\new{Likewise, we leave challenges and themes related to specific sub-fields of visualization, such as evaluation or storytelling, for another paper and instead focus on a general overview over challenges, while avoiding challenges that we deem through our discussions to be too specific or too general.}{ben, jason}

\textbf{After agreeing to the six initial challenge themes, we formed a working group for each} to further develop the themes and related challenges. All Dagstuhl participants were invited to join a group, the 21 participants who did so are authors of this paper. Tasked with \textbf{extending, refining, merging, and scrutinizing} and, finally, describing challenges within each theme, the six groups worked asynchronously for approximately two-three months, with weekly discussion meetings.

The next stage was \textbf{(4) formulating, reviewing and further refining} the challenges. A review process within and across groups
allowed us to evaluate the complexity of individual challenges and how specific these were to visualization education. At this point, we started differentiating challenges (again) into challenges for \textit{research} and challenges from \textit{practice}. We realized that our challenges would \textbf{reflect our own experience in education practice} which would make them much more grounded than general challenges. We also found that each practical challenge led to several research challenges, which we decided to \textbf{frame as research questions} from here on.
\new{Each can be preceded with a common stem, such as \say{How can we...}, and allowed adverbs to be inserted as required -- \say{effectively, efficiently, reliably, adequately, provisionally}, etc.}{jason}
This differentiation helped us focus and frame the contribution of this paper, including the opportunities in our call-to-action. For example at this point, we decided to omit generic challenges with regard to \textit{scalability} because these represented general challenges relevant but not specific to visualization. Some challenges, that were too specific were merged with others; e.g., we merged challenges on teaching children with challenges around learners more generally. Terminology and theme names were also reviewed as part of this phase. For example, we decided to rename the \textit{Audience} theme into \textit{People} to reflect an active and non-hierarchical connotation. 
This review process resulted in 59 challenges (LIST-2), distributed across six themes: \textit{People} (formerly Audience), \textit{Goals \& Assessment} (formerly Content \& Evaluation), \textit{Methods, Motivation, Resources, and Diversity (new)}. Some themes, such as \textit{People}, included as few as 4 challenges; others such as \textit{Methods} up to 13 challenges. Some of these challenges were organized hierarchically, with super and sub-challenges. 

To refine these further, we held \textbf{weekly (one hour) ``town hall’’ meetings} attended by between half and all of the authors, including at least one representative from each working group. These meetings were held over the course of four months and discussed 
\textit{(a)}~what challenges to include within each theme, and \textit{(b)}~how to  frame these consistently across the themes. In between town-hall meetings, groups were tasked to 
\textit{i)} within their theme, identify those challenges most relevant to visualization education, 
\textit{ii)} write detailed descriptions of these, and 
\textit{iii)} identify open research questions emerging from these challenges to be addressed by the community to advance our understanding and practice of visualization teaching and learning. As part of this process (LIST-3), some groups characterized challenges not captured in LIST-1 and LIST-2. Through this process, our goal was to express the challenges in an inspirational way.

\textbf{Continual discussion, reflection, revision and review} enabled us to converge towards the final set of challenges and  themes described in \cref{sec:challenges} (LIST-4), and inform opportunities in our \textbf{call-to-action}. To achieve consistency across themes and challenges, we agreed on a consistent structure for each challenge, comprising 
\textit{i)}~a headline (headed by icons in \cref{sec:challenges}),
\textit{ii)}~a text describing the challenge and limitations of current approaches, and
\textit{iii)}~one or more open research questions to be addressed as part of research and practice (concluding each challenge in \cref{sec:challenges}).
This final set of themes, challenges and questions was reviewed again across groups in the wake of \textbf{(5) formulating opportunities}.
For instance, we considered diversity an opportunity for our call to action (\odiversity), rather than a challenge. Opportunities (\cref{sec:opportunities}) resulted again from extensive discussions during the town hall meetings, which were informed by our challenges and individual group discussions. The objective of these opportunities is to find actionable pathways to address our challenges and associated research questions.
Their development resulted in a late change -- a new theme of \secchange with two challenges that took shape only while opportunities were discussed.
The paper was written in a flexible and collaborative way, creating new discussions 
and framing
a shared understanding and story.

\section{Challenges in Practice and Research Questions}
\label{sec:challenges}

\begin{table}[]
\footnotesize
    \centering
    \begin{tabular}{@{} p{.6cm}|p{7.6cm} @{} }
        \textbf{Tag} & \textbf{Challenge} \\
         \hline
         \hline
            & \textbf{\secpeople} \\
         \hline
            \chpeopleone & Impact of educators' diverse backgrounds on visualization teaching \\
            \chpeopletwo & Leveraging and catering to learners’ diverse backgrounds, goals, and needs \\

            \chpeoplethree & Acknowledging and embracing diversity to allow for mutual learning \\
         \hline
            & \textbf{\secgoals} \\
        \hline
          \chgoalsone & Identifying learning goals and designing objectives tailored to specific groups of learners \\
          \chgoalstwo & \new{Assessing creative, project-based, and problem-oriented work in a fair and efficient manner}{soren} \\
          \chgoalsthree & Assessing learners' work at scale and distance  \\
         \hline
            & \textbf{\secmotivation} \\
         \hline 
        \chmotivationone & Communicating the need \textit{for} visualization education \\          
        \chmotivationtwo & Retaining motivation \textit{during} learning visualization \\ \hline
            & \textbf{\secmethods} \\ 
         \hline
        \chmethodsone & Fostering core skills around visual representation and interaction \\
        \chmethodstwo & Developing ‘specific’ and ‘general’ skills and competencies \\
        \chmethodsthree & Adapting methods to learners and contexts \\ 
        \hline
            & \textbf{\secenvironment} \\ 
         \hline
        \chenvironmentone & Providing environments for hands-on, creative, and collaborative work\\
        \chenvironmenttwo & Using online, hybrid, informal \& workplace environments \\ 
        \hline 
            & \textbf{\secresources} \\ 
        \hline
        \chresourcesone & Finding, evaluating, and adapting materials \\
        \chresourcestwo & Reusing and adapting materials \\
        \chresourcesthree & Creating and updating materials \\
        \chresourcesfour & Creating materials for informal, self-paced learning \\
        \hline 

            & \textbf{\secchange} \\ 
        \hline
        \chchangeone & Understanding the role and effects of AI \\
        \chchangetwo & Overcoming inertia and adapting to change \\

                \hline 
        \hline 
    \end{tabular}
    \caption{Overview of our 19 challenges grouped into seven themes, all discussed in Section \ref{sec:challenges}.
    }
    \label{tab:challenges}
    \vspace{-2em}
\end{table}

This section discusses our final set of \numchallenges{} challenges, summarized in \cref{tab:challenges}. 
We start
with challenges
related to \secpeople, as people are core to education and have strong implications for any of the challenges in the other themes. \new{We then continue with themes in the order in which decisions in these themes inform each other: \secgoals{} and \secmotivation, then 
\secmethods{} and \secenvironment{} which have 
\new{a mutual}{soren}
relationship, and eventually \secresources{} and 
\new{the overarching}{soren}
\new{theme of \secchange.}{}
}{ben}

\subsection{Challenge theme: \secpeople{} (\chpeople)}
\label{sec:people}

We differentiate between \textit{educators} (those planning, preparing, conducting, facilitating, and evaluating learning), and \textit{learners} (those wanting to gain experience and skills whether obliged or voluntarily). As data visualization education spreads across disciplines, educators and learners increasingly come from a vast range of backgrounds and levels of education, industry and academia, from different segments of society, age groups, cultures, disciplines, and work environments, and have different objectives in learning about visualizations. \new{In fact, many disciplines currently discuss challenges and strategies for incorporating data analysis and data visualization in their course offerings and curricula in, e.g., Art~\cite{bertling_case_2021}, Computer Science~\cite{SpecialIssueVisLiteracy}, Design~\cite{parsons_understanding_2022, dong_aesthetic_2007}, Digital Humanities~\cite{ballentine_digital_2022}, English Studies~\cite{graham_introduction_2017}, Math~\cite{setiawan_exploring_2021}, Political Sciences~\cite{henshaw_data_2018}, Statistics~\cite{loy_supporting_2019}, Secondary Education~\cite{kahn_learning_2021, lee_data_2018}).}{isabel}

\minifig{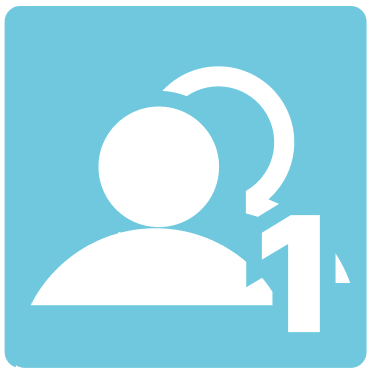}\challenge{\chpeopleone}{\new{Impact of educators’ diverse backgrounds on visualization teaching}{soren}}\new{Some educators \textbf{have training in education but many of us in academia do not}. Likewise, \textbf{disciplinary backgrounds} and \textbf{familiarity with visualization}}{ben} influence the topics educators emphasize and their methods. 
 For example, educators in the humanities might emphasize certain visualization techniques (e.g., text visualization) as well as critical and ethical perspectives on data and their visualization (e.g., processes of data collection and preparation~\cite{Drucker}, uncertainty~\cite{knox_ethnography_2018}); design educators might focus on the data representation processes and the impact of design choices~\cite{cairo_how_2019}; computer scientists and mathematicians might emphasize algorithms and statistics---or not. While this diversity accounts for richness, 
educators can feel challenged to come out of their comfort zone and respond to new topics and learner diversity (see \chpeopletwo). There seems to be little literature 
that helps prepare educators to embrace the diversity of (disciplinary) backgrounds while overcoming domain biases and knowledge gaps~\cite{brooks_towards_2021}. 
Educators would benefit from research that \new{asks:}{isabel} 
\question{\rqPPLhow}
\question{\rqPPLbck}
\question{\rqPPLrng}

\minifig{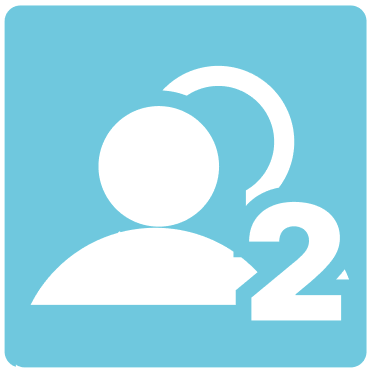}\challenge{\chpeopletwo}{Leveraging and catering to learners’ diverse backgrounds, goals, and needs}%
While learners can align in their goals and skills, they remain rich individuals and different learning needs exist~\cite{borner_investigating_2016,alper_visualization_2017,chevalier_observations_2018,bishop_construct--vis_2019}. Differences that impact learning experiences at all ages include \textbf{visible and non-visible disabilities}, such as dyslexia, which 
some of our students highlighted in relation to Lego-based constructive learning activities.
While anecdotal, it suggests a need for further research (see \cref{sec:opportunities}). \textbf{Culture} further informs and shapes visual representations from reading direction to symbolic meanings of color and images but also critical and ethical discussions about data and their representations ~\cite{correll_ethical_2019,dignazio_data_2020,schofield_indexicality_2013}. Other factors that influence topics and methods taught include \textbf{educational level}, \textbf{ability to commitment}, and \textbf{individual goals}. For example, different professional needs (e.g., uses visualization as part of their work, or simply needs help with a specific task such as designing a dashboard or using a specific tool) might invite different visualization learning experiences.
Eventually, visualization courses are increasingly characterized by learners with a range of \textbf{disciplinary backgrounds}, \textbf{skills}, and \textbf{values} which can be \textit{``astoundingly diverse''}~\cite{elmqvist2012leveraging}, such as whether someone can or wants to learn to program.

As we embark upon efforts to broaden, de-colonize, and diversify the curriculum, it is a challenge for educators to identify and select appropriate learning goals (\secgoals), methods (\secmethods), and resources (\secresources) best adapted to learners' diversity. There is an opportunity, if not obligation, to devise specific pedagogical approaches and resources for diverse, non-homogeneous groups of learners \cite{elmqvist2012leveraging}
(see \chresourcestwo; \chgoalsone), informing our efforts by asking:
\question{\rqPPLlrn}
\question{\rqPPLdiv}

\minifig{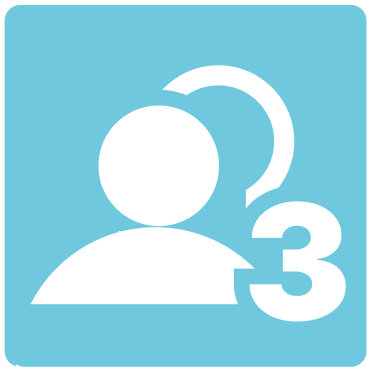}\challenge{\chpeoplethree}{Acknowledging and embracing diversity to allow for mutual learning}%
The diversites described in \chpeopleone{} and \chpeopletwo{} pose a challenge to the respective interactions between educators and learners as well as among learners themselves~\cite{adhikari_issue_2021,SpecialIssueVisLiteracy,owen2013visualization,rushmeier_revisiting_2007}. 
For example, educators need to \textbf{identify and address existing skill sets and knowledge gaps} among learners. \new{Likewise, due to the evolving and applied nature of visualization, we also have experienced cases where \textbf{individual learners contribute valuable knowledge and lessons} which can be both an opportunity and a challenge as learners and educators get pushed outside their comfort zone, change roles, and identify their respective skills and contributions to educational experiences.}{ben}
The increase of \textbf{interdisciplinary, large-scale visualization courses} requires further re-thinking our roles as visualization educators, the pedagogical strategies we apply, \new{how we \textbf{cater to learners' diversity at scale}, and enable legitimate participation~\cite{lave1991situated}. Consequently, we see a need to consider notions of communities of practice~\cite{wenger1999communities} and inquiry~\cite{garrison2016thinking}.}{soren, jason SUMMARY1} 
Research is required to establish:
\question{\rqPPLent}
\question{\rqPPLflx}
\question{\rqPPLcol}
\question{\rqPPLper}

\subsection{Challenge theme: \secgoals{} (\chgoals)}
\label{sec:learninggoals}

Learning goals are important to inform course content and set expectations about what the learner will be able to know and do~\cite{h_brown_essentials_2020}. They also inform teaching methods (\secmethods) and plan assessment against more fine-grained learning objectives~\cite{melton_objectives_2014}. Assessment, in turn, effects learning in a closely related cycle \cite{Pereira2016}; it can happen \textit{during} the process of learning (formative assessment) to help learners improve, or at the end for accreditation (summative assessment)~\cite{Biggs2011}. This section discusses three respective challenges in the context of visualization: identifying and communicating learning goals (\chgoalsone); the
fair and reliable assessment of creative, project-based, and problem-oriented work (\chgoalstwo); and fair and efficient assessment across learners and at scale (\chgoalsthree).

\minifig{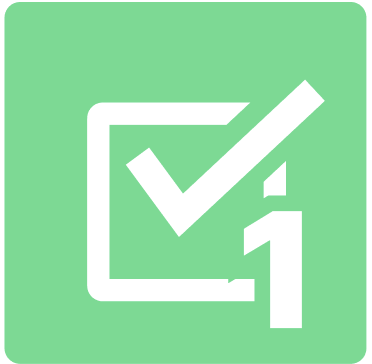}\challenge{\chgoalsone}{Identifying learning goals and designing objectives tailored to specific groups of learners}
There is currently no agreed-on list or taxonomy of learning goals for visualization nor a detailed understanding of what skills people need. 
Some learning goals relate to \textbf{theoretic visualization knowledge} (e.g., Gestalt principles, color theory, or Norman's model of interaction). Others involve the \textbf{application} of that knowledge (e.g., constructing a dashboard with Tableau) and \textbf{technical knowledge} about data, algorithms, and programming. Some require \textbf{meta-cognitive} capabilities that involve self-awareness, self-analysis, and critique; \textbf{creative and problem-solving skills} (e.g., design exploration and evaluation, making design tradeoffs, problem definition, prototyping); and again 
others may focus on \textbf{people skills} such as collaboration and communication~\cite{Biggs2011,domik_fostering_2012}.
In visualization literature \cite{Lee_Adar2023, Amabili2021, Keck2021}, Bloom's Taxonomy~\cite{bloom_taxonomy_1956} is frequently referenced to develop intended learning outcomes at different levels of complexity.
However, certain learning processes that are typical for computer science challenge this taxonomy~\cite{Johnson2006}. This especially happens in visualization, where tasks in lower levels, such as \textit{Understanding} (level 2), also involve processes normally assigned to higher levels of complexity, such as \textit{Evaluating} (level 5).
Applying the \textit{Revised Taxonomy} \cite{krathwohl2002revision}
, which captures the cognitive and knowledge dimensions of educational objectives, can help define learning goals tailored to both learners and the discipline of visualization~\cite{Keck2021}.
Mapping educational offerings to existing taxonomies is likely to reveal what educators in visualization do, how this compares, where the gaps are, and how well this theoretical model applies to the field. \new{However, educators \textbf{need to strike a balance between what they think is necessary to know about visualization and those diverse learners' (\chpeopletwo) goals}.}{ben} We need to know:
\question{\rqGAgen}
\question{\rqGAgrp}

\minifig{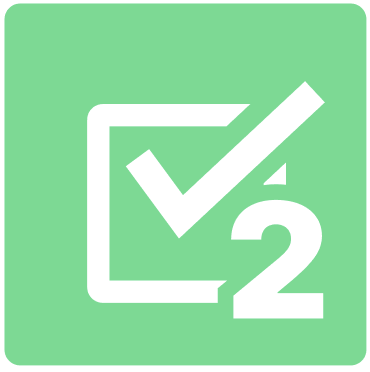}\challenge{\chgoalstwo}{Assessing creative, project-based, and problem-oriented work in a fair and efficient manner}%
Many of the goals and skills mentioned in \chgoalsone{} are hard to assess because they involve \textbf{qualitative judgment}, \textbf{factors that might be unknown to the educator}, or simply \textbf{decisions about fairness}. For example, \textbf{group work} makes it hard to assess individual contributions, especially in interdisciplinary groups where students might develop different skills. \textbf{Assessing design rationales} further requires context and many factors to be taken into account for a fair assessment.
\new{Work-based learning \cite{murtazin2020literature}, through placements and internships offers diverse applied learning experiences that must be assessed fairly and consistently.}{jason}
Biggs and Tang note that qualitative assessment is often criticized for being \say{'subjective' and 'unreliable'}~\cite{Biggs2011}.
But `subjective' does not need not be `unreliable', to which end Biggs and Tang offer strategies for subjective assessment that is transparent, fair, and reliable.

A further challenge here is \textbf{assessing learning at distance} outside traditional classroom teaching, e.g., where learners learn visualization tools and methods over an extended period of time~\cite{alkadi2022understanding} or already \textbf{bring different skills}. Fairness is an issue, especially in the case of \textbf{applied visualization problems} that involve \textbf{diverse data sets}, require students to choose \textbf{different tools} and may involve different \textbf{collaborators} and settings.
Inviting students to use their own data sets or use existing real-world data sets~\cite{Corneli2018, syeda2020design} makes a fair assessment difficult since different data sets of varying complexity may have been chosen. In cases where learners are given the freedom to choose a specific tool, fair assessment is difficult because of the different efforts required to master each tool. Educators may also lack the knowledge and experience to support learners in using and learning the wide range of tools and frameworks best suited to their task. 

The challenge is to adapt and adopt strategies to the multi-disciplinary visualization education context, but \new{little empirical evidence exists for assessment in visualization.}{ben} \new{A recent study by Beasly et al.~\cite{beasley2021through} showed that peer review in visualization classes can provide students with multiple perspectives on their project, support active learning and engagement with course content and visualization terminology, as well as inspire students through exposure to other students' work. However, the same study suggests the nature of peer review makes it best suited for formative evaluation.}{ben}
We need to rethink our reward systems in formal education if we want to address these challenges in assessment and know more about:
\question{\rqGAskl}
\question{\rqGAdiv}
\question{\rqGAdif}

\minifig{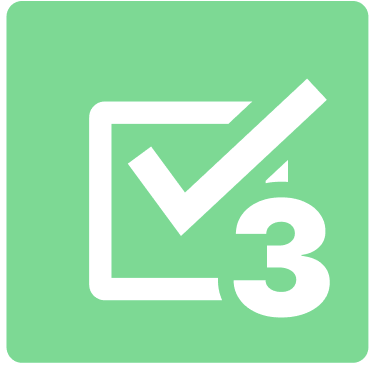}\challenge{\chgoalsthree}{\new{Assessing learners’ work at scale and distance}{soren}}
In the case of a single learner, an educator can invest time to know them and their background, help when they struggle, and assess how much they were able to achieve. In large classes, such qualitative and individual assessment is \textbf{time intense} \new{as educators read through design rationals, understand sketches, understand iterations, critique visualizations, click through interactive prototypes, and write qualitative feedback.}{ben} Formative feedback \textbf{must be returned in a timely manner} to be of value to a learner, especially in projects that require constant iteration and milestones~\cite{syeda2020design}. One option to manage marking and supervision loads is through group work or \new{peer review \cite{beasley2021through} as discussed above}{ben, jason}.
\new{Eventually, we need to provide feedback and \textbf{guidance to those learners we have no direct contact} with, e.g., those people wanting to use sophisticated visualization and visual analytics systems~\cite{alkadi2022understanding}.}{ben}
We need to ask:
\question{\rqGAfbk}

\subsection{Challenge theme: \textsc{Motivation (\chmotivation)}}
\label{sec:motivation}

Motivation is essential to effective learning at various levels \cite{schunk2014motivation}. We found many of our learners in visualization to be highly motivated, wanting to improve their analytical capability, understand design, and learn good visualization examples and practices. Despite this, our experience led us to identify challenges in motivating educational establishments to offer visualization courses (\chmotivationone) and in maintaining the motivation of those who have embarked upon such courses (\chmotivationtwo).

\minifig{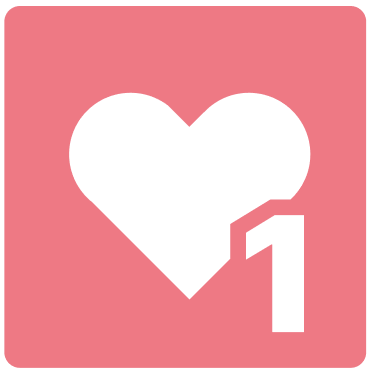}\challenge{\chmotivationone}{Communicating the need \textit{for} visualization education}%
In our experience, \textbf{people have different ideas} of what visualization is and what it means to them. This can range from making `pretty pictures' to using tools like PowerBI or Tableau, improving figures for a paper, building dashboards, designing novel visualization tools, researching visualization, or creating data art. While some people think visualization is `cool' or necessary, they \textbf{may underestimate the complexity, science, skills, and methods in visualization}. 
Yet, the notion that visualizations are intuitive and understood with little effort is not always true. 

To support this understanding, we need more persuasive real-world examples where we can \textbf{demonstrate the impact of misinformation} with loss of money, loss of time, and effect on people’s health.
Perhaps looking more broadly into other disciplines, such as Human Factors engineering and searching for use cases of data visualization, can be a potential solution. Likewise, a challenge is to \textbf{convince people, organizations (e.g., universities, industry), and policymakers} to invest time and money in visualization education and encourage people to take these courses. For example, although some aspects of data visualization literacy, such as reading graphs in mathematics or sciences, are often taught in schools, there is yet not enough formal training in visualization in schools for young learners~\cite{chevalier_observations_2018, alper_visualization_2017}.
We need research that \new{helps us understand:}{Mandy, Isabel}
\question{\rqMTVdiv}
\question{\rqMTVfac}
\question{\rqMTVsuc}

\minifig{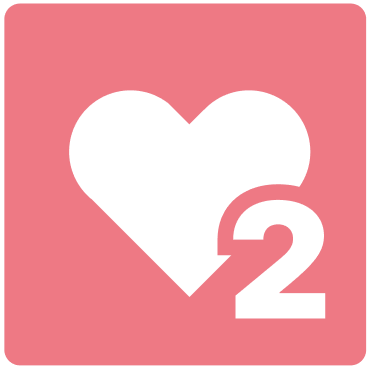}\challenge{\chmotivationtwo}{Retaining motivation \textit{during} 
learning visualization}
We found people from different backgrounds need to \textbf{catch up on knowledge and skills from the missing domains} (e.g., graphics, programming, analysis), while other learners excel in these skills (\chpeopletwo). 
\new{In our experience, this issue is most common in gaining a sense of design process and methods on the one hand and core computer science skills and approaches on the other}{soren}.
\new{While potentially demotivating, this experience can be an opportunity for learners to engage in meaningful learning communities~\cite{wenger1999communities}, where they are expected to and responsible for learning from and with each other based on their individual backgrounds.
}{soren, jason, SUMMARY1} 
In cases where a course involves project work with milestones, \textbf{specific project stages may be more difficult for some learners} if they lack the basics (e.g., programming). Group-work again comes with the risk that some learners \textbf{may feel excluded} or \textbf{`pushed' into specific tasks} (programming, designing), rather than being able to learn the skills they lack~\cite{burch2020more}. 

One solution to motivate can be to involve learners working with self-chosen real-world problems and collaborators~\cite{Corneli2018,syeda2020design} or letting learners select tools and technologies to design and create data visualizations as discussed in \chgoalstwo{} and \new{as advised by experiential learning~\cite{dewey1938experience} and problem-based learning theory}{soren}. 
\new{While we rarely expect students to arrive with a specific understanding of application domains for visualization (e.g., health sciences, natural sciences, or communication studies), such domain understanding presents additional opportunities to foster communities of inquiry~\cite{garrison2016thinking}}{soren}.
Offering a learner-centered (free-form) approach ensures that learners
create an artifact with meaning and purpose\new{~\cite[p. 67--72]{dewey1938experience}}{soren} in their research, practice, or personal life.
However,
it introduces other challenges, such as the high cost of assessment (\chgoalsthree), running courses in this format, and difficulties with tailoring the course content. Similarly, \textbf{projects may turn out to be too complex} for the time and resources of the course and \textbf{learners can easily underestimate the skills required}: data collection, cleaning, preparation, learning tools (or programming), exploring designs, and iterating. 
In particular, this can be a problem in \textbf{informal and self-regulated learning} where people don't study full-time but either learn as they go or take a course after hours in a self-paced manner. We ask: 
\vspace{-0.03cm}
\question{\rqMTVbck}
\question{\rqMTVflx}

\vspace{-0.1cm}
\subsection{Challenge theme: \textsc{Methods (\chmethods)}}
\label{sec:methods}

Common methods in visualization education include lectures, sketch-based exercises, brainstorming sessions, design critiques, co-design, project work, group discussions, game playing~\cite{adar2022roboviz}, tutorials with visualization tools and programming, peer feedback, project work and presentations, and many others.
\new{This places the activities that learners \textit{do}~\cite{goodyear2021activity} at the center of much of our education methods.}{soren}
An educator \new{then need to consider how to design~\cite{goodyear2015teaching} for this \textit{doing}, that is,}{soren} to find (or invent!) methods for teaching core skills around visual representation and interactivity (\chmethodsone), teaching soft-skills required for visualization (\chmethodstwo), and finding solutions for adapting existing methods and activities (\chmethodsthree).

\minifig{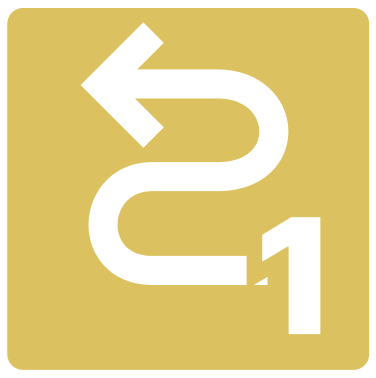}\challenge{\chmethodsone}{Fostering core skills around visual representation and interaction}%
\new{Reading, designing, and creating visual representations of data are core skills in visualization literacy~\cite{boy_principled_2014,borner_data_2019}. These skills include reading sophisticated visualization techniques and choosing among a growing number of visual representations, creating efficient visual mappings, and choosing and employing the tools to create and deploy them. These processes come with a visualization-specific body of knowledge and skills as outlined in our Introduction and that are highly specific to visualization: visual communication, visualization evaluation, theoretic models and guidelines, design patterns and processes, etc. }{ben}
 
Specific methods to teach these core skills range from general \textbf{methods embodied in tools and approaches}, e.g., learning novel visualization activities through analogy~\cite{ruchikachorn2015learning} or cheat sheets~\cite{wang2020cheatsheets}, designing visualizations by demonstration~\cite{zong2020lyra}, discussing general visualization guidelines~\cite{diehl2018visguides}, designing with card decks~\cite{he2016v} or practicing research methods in evaluation~\cite{lam2011empirical}. On the other hand, methods can include specific \textbf{hands-on visualization activities} and workshops that provide detailed steps how to engage with a given topic~\cite{huron2021ieee,kerzner2018framework,wang2019teaching,saket2016visualization}. While potentially tedious to manipulate and unfamilar to many people, \textbf{data physicalizations} have been shown to better help learn about data and its representation, compared to using user interfaces to design visualizations~\cite{wun2016comparing, huron2014constructive}. Interaction and play can be further strong factors in learning visualization (e.g.,~\cite{kiriphys}). 
 While many interfaces provide interactive exploration, discovering and \textbf{learning about the interaction} means provided by such interfaces is difficult because both interaction paradigms and people's preferences vary greatly~\cite{blascheck-discoverability,perin:thesis,saket-dm}. 
As a result, the possibilities for interaction styles and paradigms are large and often specific to factors such as data type, layout, audience, and medium, calling for a concept of interaction literacy in visualization~\cite{bach2018ceci}. We need to know: \question{\rqMTHskl}
\question{\rqMTHply}
\question{\rqMTHint}


\minifig{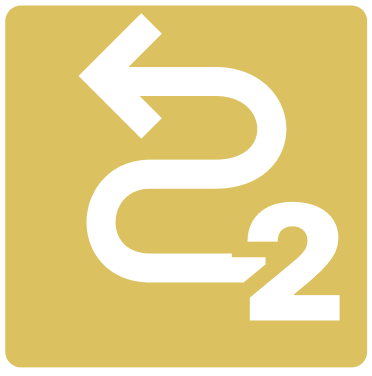}\challenge{\chmethodstwo}{Developing `specific' and `general' skills and competencies}%
In addition to the \textit{specific} focus on visual representations and interactivity (\chmethodsone), visualization requires a wide range of \textbf{skills and competencies that are more \textit{general}} in their applicability such as those discussed in \chgoalsone: problem solving, critical thinking, creativity, collaboration, group work, debate and consensus-finding, project pitching, storytelling, visual design, communication, analytic thinking.
As such, there is scope for developing these skills in visualization focussed education.
This can be achieved through group projects, self-assessment, project-based assignments~\cite{kammer_experience_2021} or collaboration with external parties, their data, and problems to add some real-world experience to visualization scenarios~\cite{Corneli2018,syeda2020design,domik_fostering_2012}. 
Alternatively, educators may focus directly on these broader skills and competencies through visualization activities that engage learners with visualization topics~\cite{byrd2021activity,beyer2021visualization,huron2020ieee} (\chmotivationtwo \& \chgoalsthree). 
Here we ask: 
\question{\rqMTHcmb}

\minifig{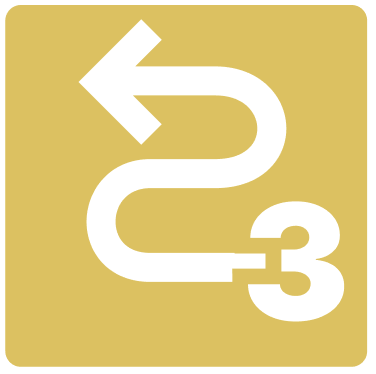}\challenge{\chmethodsthree}{Adapting methods to learners and contexts}%
Methods that work for one group of learners (\secpeople) or in one context (\secenvironment), might not work as well in another one. 
In addition, methods and activities may \textbf{not scale well with the number of learners}, may \textbf{not be possible in a given environment or context}, or a particular data set may not support certain activities. This makes \textbf{re-using methods challenging} even across instances of the same learning activity and limits innovation in developing and using methods.
We need empirical evidence to answer: 
\question{\rqMTHfnd}
\question{\rqMTHwrk}

\subsection{Challenge theme: \textsc{Learning Environments} (\chenvironment)}
\label{sec:learning environment}

As a learning environment we describe where and under which conditions learning happens. This can include in-person, online, or hybrid settings for formal learning (structured and organized by an educator) or informal (self-regulated, open, unguided); 
\new{traditional classroom based settings or work-based learning}{jason}; and can vary in number of learners, educators and their ratio.
All these factors inform methods (\secmethods) and the learning experience, interactions between learners and educators (\secpeople) and their respective social relationships~\cite{preves_classroom_2009,tygel_contributions_2016}. Here, we discuss environments in general (\chenvironmentone), as well as online and hybrid environments in particular (\chenvironmenttwo).

\minifig{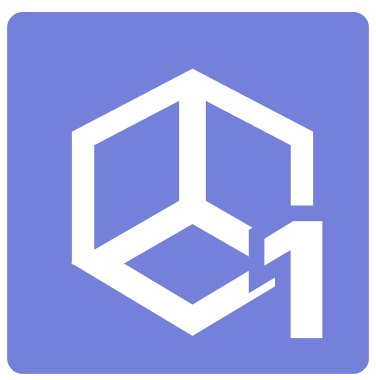}\challenge{\chenvironmentone}{Providing environments for hands-on, creative, and collaborative work}Practical and hands-on activities 
 (\secmethods) often \textbf{require dedicated spaces and equipment} such as computers, presentation and exhibition spaces, break-out rooms, printers, maker labs, etc. However, the \textbf{available resources are fixed} and cannot easily be altered. Each choice of space has an \textbf{impact on the learning experience}, for example, in an auditorium, an educator is spatially separated from the learners, possibly creating unintended barriers and hierarchy. 
This can inhibit participation and learners' ability to proactively and creatively contribute and engage~\new{\cite{lave1991situated}}{soren} in critical activities such as discussing design choices and flawed visualizations or asking specific questions.
In contrast, in a studio space, interaction between learners and educators can be integrated and activities can be dynamic, interactive, inclusive, and hence more motivating and creative~\cite{roberts_reflections_2022}.
We need to ask:
\question{\rqENVlen}
\question{\rqENVaff}

\minifig{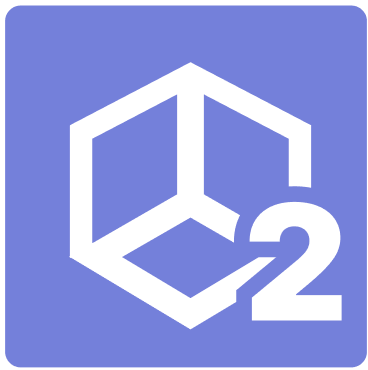}\challenge{\chenvironmenttwo}{Using online, hybrid, informal \& workplace environments}
Online teaching brings some benefits to inclusion (spatial and temporal constraints) especially, for \new{people with disabilities~\cite{bowler2022exploring} and when teaching home-learners or professionals who participate from the workplace.}{jason}
On the other hand, these environments \textbf{prohibit many of our traditional hands-on and social activities} so important in visualization and could (or not~\cite{aerts2021remote}) make them less engaging: sketching and sharing sketches, design crits, \new{group discussions, reviewing ideas and sketches.}{ben}
\new{Learning in the field, through volunteering, internships and placements has great potential in mirroring visualization research methods \cite{sedlmair2012design,hall2019design} and providing valuable and inspiring learning experiences while \say{enabling the synergy of visualization pedagogy and social good}\cite{syeda2020design}.}{jason} We have to ask:
\question{\rqENVtch}
\question{\rqENVdst}

\subsection{Challenge theme: \textsc{Materials (\chresources)}} 

As \textit{material}, we consider any artifact (or content thereof) to support education, learning, and training. This can include lecture slides, recorded video lectures, text books, tutorial outlines, curricula,\footnote{\url{http://education.siggraph.org/resources/visualization/education}} learning goals, exams and quizzes, positive and negative visualization examples, websites and blogs, guidelines \cite{diehl2018visguides}, research papers, or descriptions of visualization activities \cite{huron2020ieee}. While the number of visualization materials is steadily growing~\cite{liu:visualization}, few of them have been created for the context 
\new{of education}{soren}%
---i.e., with guidance and pedagogic principles in mind---and it is unclear which of these materials best support particular educational contexts. This seemingly paradox situation leads to a range of challenges, starting from finding and evaluating materials (\chresourcesone), to adapting them (\chresourcestwo), to creating materials and sharing them (\chresourcesthree), and creating materials for informal, self-paced learning (\chresourcesfour).

\minifig{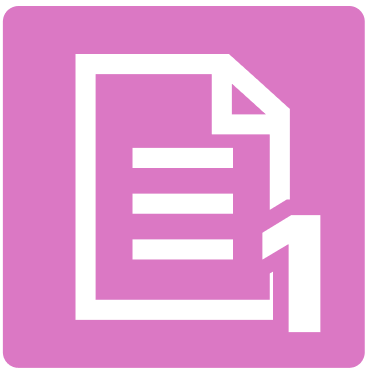}\challenge{\chresourcesone}{Finding, evaluating, and adapting materials}Materials are created by a wide range of people from practice and education. Each artifact includes different examples and angles on specific topics and offers different means of engagement (\chpeopleone). At the same time \textbf{credibility of materials} can vary from established media (books, peer-reviewed papers) to tools and posts on the web. Likewise, material needs to \textbf{specifically support learners, methods and the environment} of a course. While experienced visualization educators might be more knowledgeable in the landscape of visualization materials, \textbf{novice-educators} and learners might adopt a material they first found or that was suggested by a colleague. \new{On the other side, \textbf{experienced educators might lack incentive to update their materials and methods}.}{ben}
Current solutions to finding and evaluating materials include investing \textbf{hours of time studying individual materials} and \textbf{regularly searching for updates}.
This leads to questions: 
\question{\rqMATfnd}
\question{\rqMATwrk}

\minifig{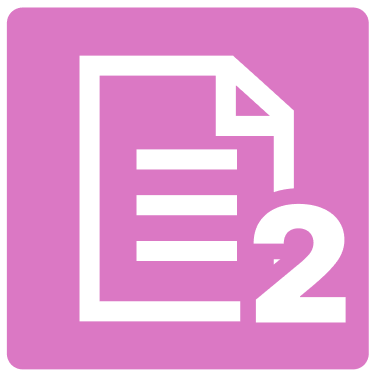}\challenge{\chresourcestwo}{Reusing and adapting materials}Even if there are many educational resources, they may need to be customized to specific teaching environments, require content update or include annotations by educators. For example, an educator might want to highlight specific content, ask their learners a question, include additional references, explanations, or examples. However, some formats and artefacts are more \textbf{difficult to adapt} than others. e.g., videos. Visualization activities \textbf{might require entire redesign} to adapt them to specific learners and time budgets. On the technical level, adapting teaching materials also requires \textbf{having access to editable versions}. 

\textbf{Accessibility of learning material can be limited} through the type of media (e.g., online video, hard-copy books, interactive applications, virtual reality), their legal status, or human factors. While many visualization resources are entirely free or available through education packs, e.g., tools like RAWGraphs\footnote{\url{https://www.rawgraphs.io}}, others are only available as demo versions and otherwise involve a charge, which can be prohibitive. 
Likewise, some materials are by definition \textbf{inaccessible to people with deficiencies, disabilities, or technological barriers}, such as reliable internet, specific hardware, voice-only videos, colored visualizations, or any visual content in general being inaccessible to the blind. Eventually, we need to consider aspects of \textbf{waste and environmental sustainability} in (re)using educational materials. 
Besides paper copies and internet streaming, this is particular imminent in activities involving constructive visualization \cite{huron2014constructive} or physicalization workshops \cite{huron2017let} that demand 3D-printed tokens, paper, or wooden objects and which often may be discarded after an activity. We need research to address questions around this re-use: 
\question{\rqMATshr}
\question{\rqMATsus}
\question{\rqMATfrm}

\minifig{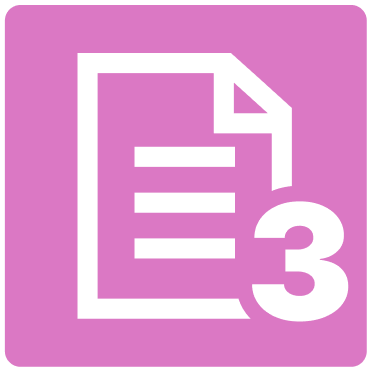}\challenge{\chresourcesthree}{Creating and updating materials}When materials are unavailable or unsuitable an educator needs to create their own material. Or, as a rapidly changing field~\cite{owen2013visualization} there simply are \textbf{no materials available for novel topics}, e.g., visualization and AI, responsive visualization, or visualization in immersive and situated environments. The lack of materials makes it harder for these topics to propagate into courses and activities. 
While \textbf{incredibly time-intensive}, creating one's own educational materials can be extremely rewarding and beneficial to both educators and students. Teachers learn through doing, too. Bespoke materials offer the most flexibility to adapt to a given learning scenario and will provide the educator a
portfolio of artifacts
to reuse and adapt in their continuing practice. 
However, visualization can \textbf{require a lot of interactive, visual, and active materials}: visualization examples, schema, code, data sets, interactive applications and tools, and activities that engage with design thinking and critical processes. This in turn can \textbf{require specific software and equipment} (e.g., for high-quality video recording), skills and time. 
Research questions can inform our efforts:
\question{\rqMATown}
\question{\rqMATgap}

\minifig{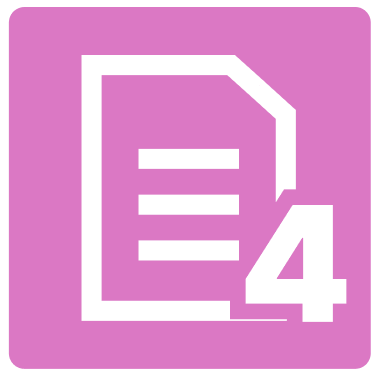}\challenge{\chresourcesfour}{Creating materials for informal, self-paced learning}Worksheets, guidelines, cheatsheets, videos, and textbooks are good materials for informal and self-paced learning but specific \textbf{materials that help with specific tasks}, \textbf{decision-making} and \textbf{iteration} are rare. 
Examples of learning visualization by doing, through engagement with personally relevant data sets and supporting people in their daily visualization tasks, and materials that support them ~\cite{perin2021students} are few. The guidelines that do exist are hard to apply in individual settings through self-paced learning. Perhaps, technology that can make recommendations in bespoke contexts and provide feedback on progress offers some potential here. For example, a tool may recommend a color scheme or interaction, respond to visualization design choices~\cite{mcnutt2018linting} and perhaps point to additional learning materials as required. 
But we need research to support efforts:  
\question{\rqMATint}

\subsection{Challenge theme: \textsc{Change ({\chchange})}}
\label{sec:change}

Developments in technology and knowledge in visualization are often sudden and variously bringing new opportunities as well as challenges. \new{Recent examples include the availability of large language models as well as leaps in the development of generative AI in general~\cite{wood2022beyond}; but also include innovation around display and interaction devices for immersive environments~\cite{ens2021grand} and portable devices; novel visualization guidelines~\cite{kay2016ish}, novel tools, or novel educational methods~\cite{syeda2020design}.}{jason} On the other side, some of what we know about visualization and ways that we use it become less valuable or even obsolete as technology (especially AI) is increasingly applied to activities that have previously required human effort. Given the time and resources that go into acquiring skills and knowledge, as well as designing materials and methods to support this learning, change is a significant challenge to many educators. 
This section discusses specific challenges related to AI (\chchangeone), and adapting and overcoming inertia to change (\chchangetwo).

\begin{wrapfigure}{l}{.6cm}
\vspace{0cm}
\includegraphics[width=.8cm]{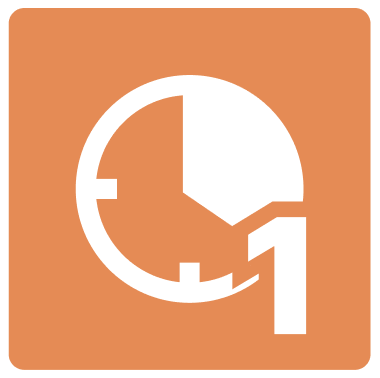}
\vspace{-0.85cm}
\end{wrapfigure}
\challenge{\chchangeone}{Understanding the role and effects of AI}%
Openly accessible AI is at the forefront of the thoughts of many at present.
It can interpret and create data, code, and graphics, and perform highly credible analysis on large data sets.
This is \textit{enabling}, offering exciting possibilities for visualization and analytics,
most obviously by building upon research in which increasingly sophisticated recommender systems capture domain knowledge
 (e.g., \cite{mackinlay2007show, moritz2018formalizing, lin2020dziban, bao2022recommendations}) that can help with \chresourcesfour.  
But the tools that are emerging are able to code, synthesize, design, analyse and critique in increasingly sophisticated and useful ways.
This is likely to have disruptive effects on visualization, how it is used, and how educators support this use, that are far wider than influencing design decisions.
This leads to a series of related questions with increasingly deep connotations that challenge our approaches and assumptions to their foundations. We ask:
\question{\rqCHGass}
\question{\rqCHGuse}
\question{\rqCHGcmp}

\minifig{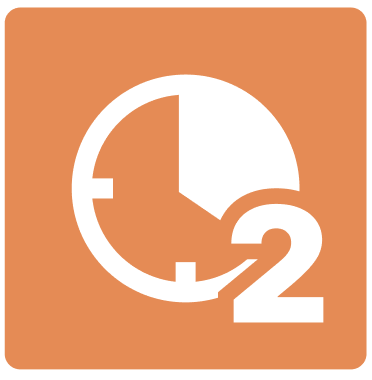}
\challenge{\chchangetwo}{Overcoming inertia and adapting to change}AI is unlikely to be the only change that impacts visualization and its education. Actively \textbf{identifying changes in technology and society} to which educators must respond is quite a challenge, as is doing so appropriately, in a timely fashion and in light of other constraints---particularly time. We need research to understand:
\question{\rqCHGfut}
\question{\rqCHGrel}

\section{Opportunities and Call to Action}
\label{sec:opportunities}

The comprehensive set of challenges related to 
data visualization education and the individual research questions (Q1-Q\numquestions) give some insights into the current state of visualization education and its open issues. 
Our seven themes---\secpeople, \secgoals, \secmotivation, \secmethods, \secenvironment, \secresources, and \secchange---should be seen as lenses, rather than taxonomic groups: the themes reflect factors important to planning, understanding, and evaluating learning, while leaving room for overlap and prioritization. \new{While many of these challenges occur beyond visualization---e.g., learner's diversity (\chpeopletwo), assessing creative and problem-solving skills (\chgoalstwo), or challenges in creating material (\chresourcesthree)---we deemed these challenges important in visualization and describe how they unfold in the specific context of data visualization. Other challenges---such as motivating the need for visualization (\chmotivationone) and teaching visualization core skills (\chmethodsone) are more specific to our domain.}{ben}
In the remainder of this section, we use these challenges and research questions to discuss how we move forward from here to create better education in visualization.

\subsection*{O1: Embrace DIVERSITY AND INCLUSION}
\label{O1:diversity}

\minifig{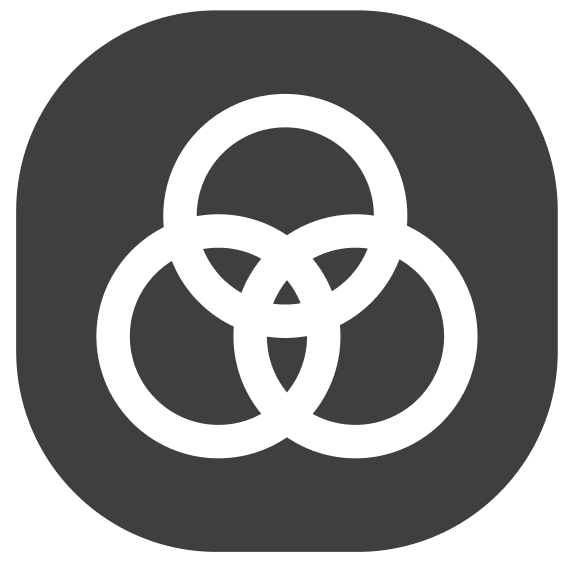}\textbf{Embrace diversity, \new{equity and accessibility}{isabel} in visualization education to develop diverse knowledge that can be applied to diverse problems \new{by diverse people (educators and learners).}{isabel}}---Diversity has been identified as a recurring thread across all challenges; from the diversity of angles on visualization education, the diversity of learners (PPL2) and educators (PPL1), to the diversity of material (MAT2, MAT3, MAT4) and methods (MTH1, MTH2, MTH3). \new{In these contexts, we had discussed diversity mainly as a challenge, however, \textbf{embracing diversity and inclusion are imperative to include the multiple perspectives, topics, ideas, and applied problems within the topic of visualization}---especially if visualization wants to be a discipline of exposing and ``making things visible'', scrutinizable, and understandable in an objective, accessible, and transparent manner. We want to celebrate and support opportunities for research across different disciplines, perspectives, and backgrounds~\cite{losev:embracing}. This could offer educators critical and equitable methods to reconsider and adopt cross-disciplinary pedagogical strategies, critical conversations and questions about diversity, inclusion, and accessibility in the future of data visualization teaching and learning praxis.}{ben}

While there is some diversity among us authors and our students (e.g. culture, gender, backgrounds), our experiences and perspectives are grounded mostly in academic settings (undergraduate, postgraduate) with limited experience working, e.g., with children, older adults, or people with disabilities. 
Our challenges are also written from the perspective of educators, especially in academia, missing experiences from educators outside of academia. Eventually, our group of authors represents a mostly Western perspective on visualization and education, while many of our learners are from other regions. That means that we are inevitably missing perspectives that are essential to this discussion. 

\new{In a recent study, Firat and Laramee~\cite{firat:inclusivity} review gender differences in spatial cognition and suggest measures to \textbf{facilitate an inclusive visualization classroom}: clarity of both oral and written content, opportunity for supplementary discussion, clarity of assessment and feedback process, discussion of common visualization examples, individually tailored data collection and visualization projects, incorporating a range of interaction and visualization technologies.}{soren} \new{However, in addition to \textbf{taking into account demographic diversity} (i.e., age, gender, ethnic and socio-economic background), we need to consider other facets of diversity: \textbf{cognitive diversity} (e.g., consideration of different knowledge structures such as different learning strategies and cognitive styles), \textbf{disciplinary diversity} (e.g., consideration and appreciation of various skill sets and a dynamic exchange of different disciplinary perspectives) and \textbf{functional diversity} (e.g., consideration of different roles students take on campus such as teaching assistant or roles they assume on group projects such as design instead of programming)~\cite{gaisch_diversity_2019}). Our call-to-action invites consideration of all these factors as we seek to design meaningful and accountable learning experiences and inclusive and accessible visualization artifacts. }{Isabel and Mandy} 

\new{We see many opportunities for research, such as \textbf{developing theoretical frameworks specifically aimed at visualization education across disciplines~\cite{losev:embracing}}. Other opportunities include, for example, \textbf{co-designing teaching and learning environments} in data visualization, co-teaching, and \textbf{developing long-term relationships with leaders and experts in equity, diversity, inclusion and accessibility} across communities of practice. These opportunities and partnerships may deepen and broaden the scope of this developing work in data 
\new{visualization classrooms and laboratories.}{soren}
Such collaborations could also inform the design of equitable data collection methods to get feedback from our learners about equity, diversity, and inclusion initiatives in data visualization classrooms. Our preliminary survey in Sec. 3.1 could be used to better understand current states and practices in visualization education beyond the set authors of this paper.}{tatiana}

\subsection*{O2: Build COMMUNITIES}
\label{O4:community}
\minifig{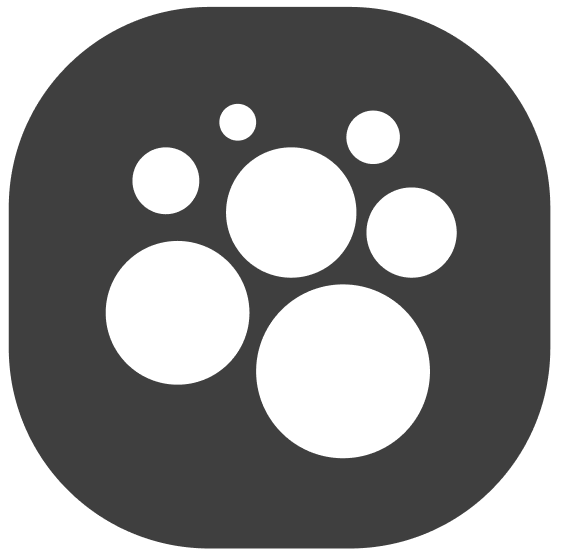}\gc{Build interdisciplinary communities for exchange and dissemination}---%
To promote diversity and inclusion and to better understand which approaches work, as well as to help disseminate practices in visualization education, we need to foster interactions between disciplines, practices, and communities. \textbf{Connecting to networks of practitioners} could help engage with people and organizations in practice to understand their approaches to visualization as well as their need for education. From our own experience we know that there is a huge potential for continuous professional development (CPD) with companies and working professionals. Examples of such networks include the DataViz Society,\footnote{\url{https://www.datavisualizationsociety.org}}  Graphic Hunters,\footnote{\url{https://graphichunters.nl}} or a wide range of local data visualization meetups. In fact, some of these communities already provide resources and activities around education in the widest sense, such as tool collections (DataViz Society) and invited public lectures (Graphic Hunters). 

\textbf{Connecting to educators and policy makers} in education beyond academia could help understand teaching and to inform visualization education on a societal level, i.e., within the landscape of school education and CPD (\chmotivationone). Eventually, it is imperative to \textbf{engage with the academic education communities} to cross-fertilize research and adopt accepted research methods (O3). 

To that end, \textbf{we need academic and non-academic, physical and online platforms} to exchange knowledge, experience, evidence, materials, activities, and guidelines. For example, the Eurographics conference has a dedicated education paper track, the Design Research community has a call for \textit{Futures in Design Education} by the \textit{Education Special Interest Group}\footnote{\url{https://shorturl.at/gjmR8}}
and in 2023 both Information+\footnote{\url{https://informationplusconference.com/2023}} and the IEEE VIS \textit{1st EduVis workshop}~\cite{eduvis2023} 
explicitly invite submissions on visualization education.

\subsection*{O3: Conduct RESEARCH}
\label{O3:research}
\minifig{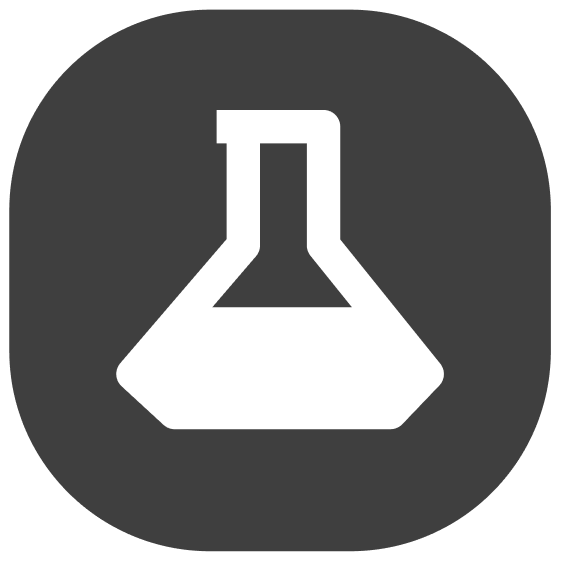}\gc{Provide actionable guidance \new{ and theory}{ben} for all educators through evidence-based research and reflection}---\new{Research is core to address challenges in visualization education and this paper has surfaced several research questions to inform theories, hypotheses, and creative solutions.}{ben} 
\new{We need \textbf{reliable empirical evidence} about the effectiveness of our approaches, reported in ways that are open to scrutiny and enable close and conceptual replication \cite{brandt2014replication}, in order to understand what works, how much and under which circumstances. Empirical evidence will help educators make better decisions in their education planning. This will be helpful to those new to education or visualization and without education training. Such a framework should \textbf{provide guidelines} and possible solutions (education design patterns?) for planning and conducting educational activities starting from specific visualization activities, to courses and curricula, \new{as well as help \textbf{formulate higher-level theories of visualization education}.}{ben} We must also \textbf{embark on ``theory borrowing''} from related educational fields such as computer science, design, or human-computer interaction. For example, to inform the design of education, we could start asking questions such as \textit{Who}, \textit{What}, \textit{Why}, \textit{When}, and \textit{Where}. Such an approach might produce the kind of pragmatic mid-level models for data visualization education that are advocated by Tedre and Pajunen in their work on the role of theory in computing education research~\cite{tedre2022grand}.}{jason}

Now, how do we plan and \textbf{conduct rigorous research} in visualization education? While some of the research questions mentioned in this paper could be solved with methods from human-computer interaction and user-centered research, others require \textbf{methods from the fields of psychology, pedagogy and education}. This includes \textbf{addressing ethical issues in research} around working with potentially vulnerable populations such as children, elderly, or disabled persons as well as testing different and replicable conditions in educational settings~\cite{brandt2014replication}. Papers that report on empirical studies in visualization education provide excellent starting points (e.g., \cite{beasley2021through}) but as is typical for human-centered research, it will take a long time and \textbf{many studies involving a diverse range of people and settings} to provide actionable evidence on the impact of specific teaching approaches.

\subsection*{O4: Act AGILE}%

\minifig{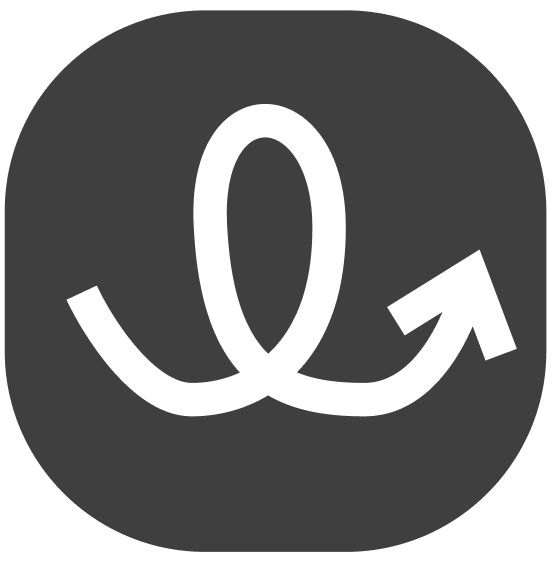}\gc{Establish mechanisms for future-thinking 
and ensuring that visualization education is dynamic, adaptable and relevant in light of changes in society and technology}---%
Ultimately, the educational community, and particularly the parts of it that are closest to the research community, \textbf{needs to ensure that its outcomes, curricula, and methods for learning and assessment remain fit for purpose} in an increasingly dynamic world (\chchangetwo), \new{in other words, that we educate people in the skills that will remain useful to them in any possible future}{ben}.
But, given our investment in knowledge, skills and education, do we have the means to detect and address the need for change?
As the technological and societal landscapes shifts, the educational landscape must adapt with it to ensure that it develops effectively, is fit for purpose (and people) and is not held back by social or technological
inertia (\chchangetwo).
This will require
\textbf{concerted collaborative efforts to conduct and share research} and transfer this into educational offerings, 
\textbf{future thinking} (such as that on the relationship between visualization and AI~\cite{wood2022beyond}), 
\textbf{engagement} with those who can identify the needs for skills and knowledge and embrace change, \new{and
\textbf{flexibility} at all levels---in courses, curricula, classes, and from all parties---institutions, accrediting bodies, educators, and learners.
}{jason}
Important opportunities and challenges for research and education will emerge, and
\ocommunity{} and \oresearch{} development will have important roles:
as a community, we will need to \textbf{think deeply, act swiftly, and research effectively} if we are to continue to empower learners and use visualization to do so.

\subsection*{O5: Relish RESPONSIBILITY} 
\label{O5:responsibility}

\minifig{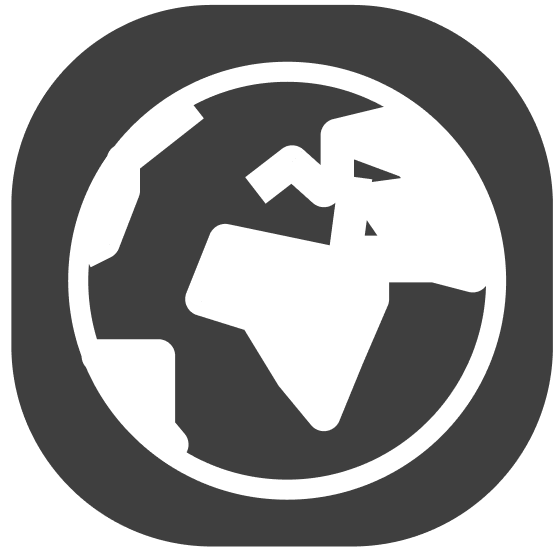}\gc{Acknowledge the impact of education on the future and take responsibility to shape this future}---All authors of this paper are passionate visualization educators, and our dedication to furthering this topic is evident in this paper and the twelve months of dedicated discussions that preceded it. However, as academics, we have to balance our time between multiple demanding activities, including teaching, research, and administrative duties. From this perspective, teaching can feel like a chore, and time constraints often make us fall back to teaching visualization in the ways we are used to, rather than allowing time for reflection and re-designing our courses (\chresourcesthree), methods of teaching and assessment, and teaching materials accordingly (\chchangetwo). 

However, we want to end with a reminder to ourselves \new{and our readers}{jason} to \textbf{understand teaching as a privilege and a responsibility}; a responsibility that goes beyond learners' individual futures, but responsibility for the future of our discipline---visualization; 
education not only embodies the knowledge, methods, and culture of a discipline and presents these to the world, but it influences the skills, principles, and values that both learners and educators will use to shape the future of that discipline: many of us have experienced that teaching and engaging with learners is an extremely gratifying experience that can be an inspiration to advance our own knowledge and skills (\chpeoplethree), reflect on our personal values, help define our discipline, and inform how these values are manifested in our research and teaching~\cite{meifestos2022} in ways that may develop into new (research) ideas.
\new{Such sentiments are, for example, voiced in notions of communities of practice~\cite{lave1991situated} and inquiry~\cite[p. 53--66]{garrison2016thinking}, which can serve as inspiration for both course design, and more broadly, in nurturing learning environments, such as research labs and other off- and online learning communities.}{soren}
Therefore, we need to \textbf{promote stronger links between research and education-related discussions} and make the topic of education and related contributions visible, in particular, at academic venues (see \textbf{O3: Community}).
\section{Conclusion}

The \numchallenges{} challenges, \numquestions{} research questions and \numgrandchallenges{} opportunities in this article, aim to map current challenges in visualization education, to expose visualization education as an emerging field, to share perspectives, to stimulate discussion, and to help reflect on visualization as both a practice and research field. This article is a \textit{call-to-action} without intending to prescribe solutions in a highly dynamic and diverse field. Rather, it aims to celebrate the richness of the field of visualization \new{education}{soren} by claiming the space and providing opportunities for research to address these challenges. We want to learn with, and from, the diverse perspectives of learners and educators in our multidisciplinary field and hope this will help foster a culture for visualization, an inclusive community, and a framework for rationalizing decisions for the many ways in which education creates interactions between the discipline, people, and the future.

\acknowledgments{%

This work was supported by \textit{Dagstuhl 22261} on \textit{Visualization Empowerment}.
We gratefully thank the organisers, all participants, and funders of the seminar for their inspiration, contributions and encouragement. 
This research was supported in part by the EPSRC (EP/S010238/2, EP/V010662/1) and NSERC RGPIN-2021-04222.
}

\newpage
\bibliographystyle{abbrv-doi-hyperref-narrow}

\bibliography{main}










\end{document}